\documentstyle[pra,aps]{revtex}
\begin{document}
\draft
\title{Supercooling across first-order phase transitions in vortex matter}
\author{P. Chaddah and S. B. Roy}
\address{Low Temperature Physics Lab\\
Centre for Advanced Technology,\\
Indore 452013, India}
\date{\today}

\begin{abstract}
Hysteresis in cycling through first-order phase transitions in vortex matter, 
akin to the well-studied phenomenon of supercooling of water, has been 
discussed in literature. Hysteresis can be seen while varying either 
temperature T or magnetic field H (and thus the density of vortices). 
Our recent work on phase transitions with two control variables
shows that the observable region of metastability of the
supercooled phase would depend on the path followed in H-T
space, and will be larger when T is lowered at constant H compared to the case
when H is lowered at constant T. We discuss the effect of
isothermal field variations on metastable supercooled states 
produced by field-cooling. This path dependence is not {\it a priori}
 applicable to metastability caused by reduced diffusivity or
 hindered kinetics. 

 Keywords Supercooling, metastability, superconductors, vortex matter,
first-order transition.
\end{abstract}
\pacs{64.60 My; 64.70 Dv; 74.60.-w}
\maketitle
In recent years first-order phase transitions in vortex matter have
been studied with both temperature and magnetic field (or vortex
density) as the control variable, and the question of
metastability has been addressed \cite{1,2,3}. The phase
 transition temperature T$_C$(H) \cite{4} drops as magnetic field is
 raised, as depicted in Fig.1. Vortex matter contracts on being
 heated from the ordered (solid) phase to the disordered (liquid)
 phase, similar
 to the behaviour of ice at pressures below 200 MPa \cite{5}.
 Hysteresis has been reported,
 with both field and temperature as the control variable,
  across the vortex-lattice-melting transition \cite{1,2}. 
We have also reported supercooling of the higher entropy 
vortex-solid phase in the polycrystalline samples of C15
 Laves phase superconductor 
CeRu$_2$, both on reducing field and on 
reducing temperature, and have found that the supercooled 
state persists farther in the latter case \cite{3,6}. Similar
signatures of supercooling have been reported in single crystals
of CeRu$_2$, NbSe$_2$ and YBa$_2$Cu$_3$O$_7$ \cite{7,8,9}.

The standard treatment \cite{10} of
supercooling across a first-order transition considers that only 
temperature is varied and other possible control variables 
(like magnetic field) 
are held constant. The free-energy density is expressed in terms of the order
parameter S as
\begin{equation}
f(T,S) = (r/2)S^2   -  wS^3   + uS^4
\end{equation}          

where w and u are positive and temperature-independent \cite{10}.
(We will assume in this paper that symmetry does not prohibit
terms of odd order. If it does, then the free energy would be
expressed as f = (r/2)S$^2$  -  wS$^4$  +  uS$^6$  ,and it is easy to 
follow and carry through all arguments in this paper. 
The assumption of the form of equation (1) is thus made without
loss of generality.) At T = T$_C$ the two stable states with f = 0 , are 
at S = 0 and at S = S$_C$ = w/(2u). These are separated by an
energy barrier peaking at S = S$_B$ = w/(4u) , 
of height f$_B$ = w$^4/$(256 u$^3$ ).
These results are independent of any assumption about the
detailed 
temperature dependence of r(T). The standard treatment \cite{10}
assumes that r (T) = a [ T - T$^*$] , where a is positive
and temperature independent, and where d$^2$f/dS$^2$ at S=0 
vanishes at T= T$^*$.
Simple algebra shows that the limit of metastability on cooling is
reached at T$^*$ = T$_C$ - w$^2/$(2ua) . 
The limit of 
metastability on heating is reached 
when the ordered state no longer has a local minimum in f(S). This occurs 
at  T$^{**}$ = T$_C$ + w$^2/$(16ua).
As noted above,  supercooling (or superheating) can persist 
till T$^*$ (or T$^{**}$)
only in the limit of infinitesimal fluctuations. The barrier
height around S=0 drops continuously as T is lowered below
T$_C$, and this is depicted in Fig.2. In the 
presence of a fluctuation of energy e$_f$, supercooling will
terminate at T$_0$ where the energy barrier satisfies
\begin{equation}
f_B(T_0) \approx [e_f + k_BT_0] 
\end{equation}
Similarly, the barrier height around the ordered state drops 
continuously to zero as T is raised towards T$^{**}$, 
and this is depicted in Fig.3. The fluctuation energy in the ordered state 
will dictate when superheating will terminate.

The formulation  stated above is of course valid for a first order 
transition in vortex-matter as a function of T. 

Vortex-matter phase transitions are encountered in H-T space and
the limit of supercooling (T$^*$) is now a function of H. This
standard treatment has recently been extended to the case where
one has two control variables, viz. density and temperature. It
has been shown \cite{11} that when T$_C$ falls with rising
density (as in water-ice below 200 MPa), then T$_C$-T$^*$ will
rise with rising density. If, on the other hand, T$_C$ rises
with rising density (as in water-ice above 200 MPa), then
T$_C$-T$^*$ will fall with rising density. This appears
consistent with experiments on ice (see Fig.5 of \cite{5}). The
density of vortices rises with increasing H, and these
predictions are also consistent with our data on single crystal
CeRu$_2$ \cite{12}. The first order phase boundary T$_C(H)$ can be
crossed by following arbitrary paths in H-T space.
It has  been argued, however, that the very
procedure of varying H introduces fluctuations, 
so that the disordered phase can be 
supercooled up to T$^*(H)$ only if T is lowered in 
constant H (i.e. in the field-cooled mode); 
fluctuations will terminate supercooling at a line T$_0(H)$
which lies above the T$^*$(H) line \cite{11}. 
For the case where H is lowered isothermally to cross T$_C$(H),
it has been argued \cite{11} that the T$_0$(H) line will be such
that T$_0$(H)-T$^*$(H) rises with rising H.

In this paper, we wish to consider the case where the sample is
cooled in constant H to a temperature T satisfying T$^*$(H)$<$T 
$<$T$^0$(H), and then subjected to an isothermal
field variation. The isothermal field variation $\Delta$H produces
a fluctuation energy e$_f$ which increases monotonically
(but nonlinearly) with $\Delta$H \cite{11}. The field-cooled state
at T corresponds to supercooling of the disordered phase, and it
sits in a local minimum of free-energy as depicted in Fig.2(b)
and 2(c). The fluctuation energy e$_f$ (and thus the isothermal
field variations, $\Delta$H) required to cross-over to the absolute
minimum in free energy (the ordered state) is clearly less at
T$_2$ than at T$_1$ because T$_2$ is closer to T$^*$ and the
free energy barrier f$_B$ defining the local minimum is smaller.
Similarly one can field-cool to (H$_1$,T$_2$) and
(H$_2$,T$_2$), with H$_2$ less than H$_1$ and T$_2$(H$_2$)
greater than T$^*$(H$_2$) as depicted schemetically in Fig.1.
Since the point (H$_2$,T$_2$) lies closer to T$^*$(H) line, it
follows immediately that the free energy barrier f$_B$
separating the disordered metastable state from the globally
ordered state will be smaller at (H$_2$,T$_2$) than at
(H$_1$,T$_2$). The fluctuation energy e$_f$, and thus the isothermal
field variation, $\Delta$H, required to cause the metastable state
to transform to the ordered sate will be smaller at H$_2$ than
at H$_1$. 
Our heuristic arguments can
similarly be used for various other experimentally accessible
paths in (H,T) space.

We have argued that an extension \cite{11} of the standard
theory of supercooling can explain various path-dependent
history effects seen in first-order phase transitions in vortex
matter. We recognize, however, that metastability does not
necessarily imply a first order transition (a sudden change in an
equilibrium property like volume does).
Metastability is also associated with glassy systems
which have been "frozen out of equilibrium". We contrast these
two types of metastabilities by considering a liquid-solid
transition. 
A liquid can be supercooled below its melting point T$_m$
without any sudden change in its diffusivity. A glass, on the
other hand, is obtained by cooling a liquid very rapidly. The
atoms freeze into an amorphous structure corresponding to the
liquid. A glass is thus a supercooled liquid that has been
frozen out of equilibrium \cite{13}. Its diffusivity dropped (and viscosity
rose to a value greater than 10$^{13}$ poise) before the atoms
could rearrange themselves from the structure of a liquid to
that of a crystal.

A supercooled liquid is thus distinct from a glass in that its
diffusivity is large enough to permit it to explore
configuration space on laboratory timescales, such that the
ergodic hypothesis is valid. Entropy is a valid concept, free
energy can be defined, and a supercooled liquid is in a local
minimum of free energy. A glass, on the other hand, explores
only its immediate
neighbourhood in configuration space, is non-ergodic, and is in a
local minimum only of the energy landscape \cite{13} and not of
the free energy. A glass is characterised by low diffusivity and
hindered kinetics, whereas a supercooled liquid is understood
by conventional statistical mechanics. Finally, it is not
necessary that a glass be associated with an underlying 
thermodynamic singularity (i.e., a first order phase
transition) \cite{14}.

We have in this paper made definite predictions on the path
dependences when metastablities are associated with supercooling
across a first order transition. These predictions do not, {\it a
priori}, apply to metastabilities caused by hindered kinetics
associated with a glass. In fact, naive arguments using hindered
kinetics suggest that since the extent of vortex motion is more
in isothermal field scans \cite{15}, this path would show more
persistent metastabilities than the field-cooled path.
Supercooling in the case of isothermal field reduction would
then persist farther than in the case of cooling in constant
field, contrary to our predictions \cite{11} for supercooling
across first order transitions. The path-dependence of 
metastabilities associated with hindered
kinetics thus needs to be quantified, beyond these naive arguments.

We gratefully acknowledge helpful discussions with Dr. S. M.
Sharma, Dr. S. K. Sikka, Dr. Srikanth Sastry, Prof. Deepak Dhar 
and Dr. Sujeet Chaudhary.

\begin{figure}
\caption{We show a schematic of the phase transition line 
T$_C$(H) and the stability limit T$^*$(H) for the supercooled
state : (H$_1$,T$_1$) and (H$_1$,T$_2$) indicate supercooled
states when vortex matter is cooled in a field H$_1$. See text
for details.}
\end{figure}

\begin{figure} 
\caption{We show schematic free energy curves for (a) T=T$_C$, (b)
T=T$_1<$T$_C$,  (c) T=T$_2<$ T$_1$ and (d) T=T$^*$. The disordered
state sits in a local minimum and is stable against
infinitesimal fluctuations for T $>$ T$^*$. This local minimum
becomes shallower as T is lowered below T$_C$ and the disordered
state at T$_2$ is unstable to a smaller fluctuation energy than
at T$_1$.}
\end{figure}
\begin{figure}
\caption{We show schematic free energy curves for (a)
T=T$_3 >$ T$_C$, (b) T=T$_4 >$ T$_3$ and (c) T=T$^{**}$. The
ordered state sits in a local minimum and is stable against
infinitesimal fluctuations for T$<$T$^{**}$.}
\end{figure}

\begin{references}
\bibitem{4}T$_C$ refers to a phase transition within vortex
matter, and not to the superconducting to normal transition.
This can correspond to vortex-lattice melting \cite{1,2}, or 
to a transition between two states with pinning \cite{3}.
\end{references}
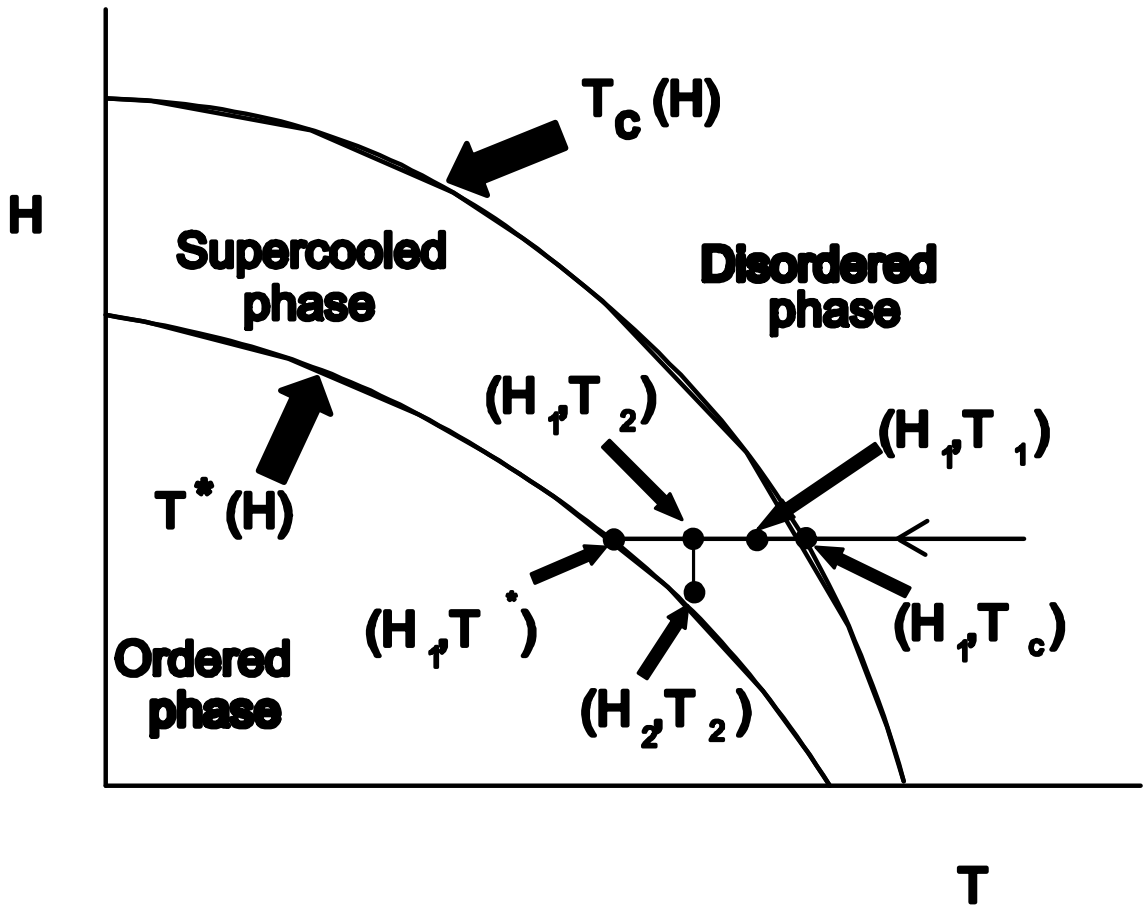
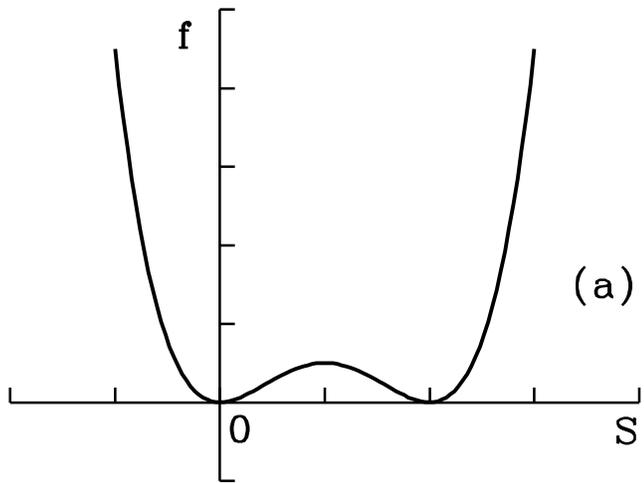
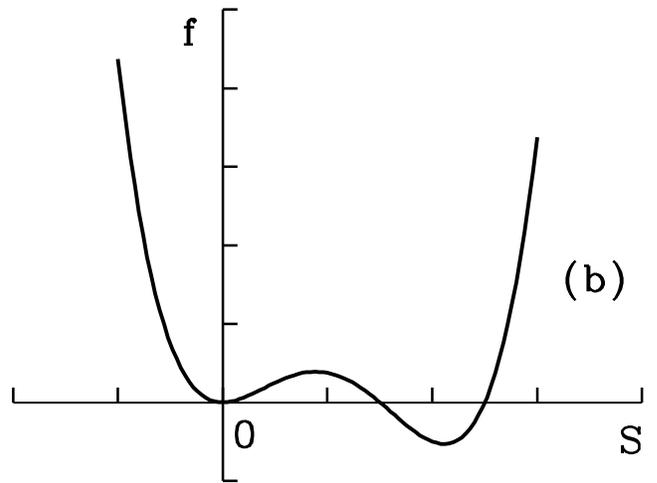
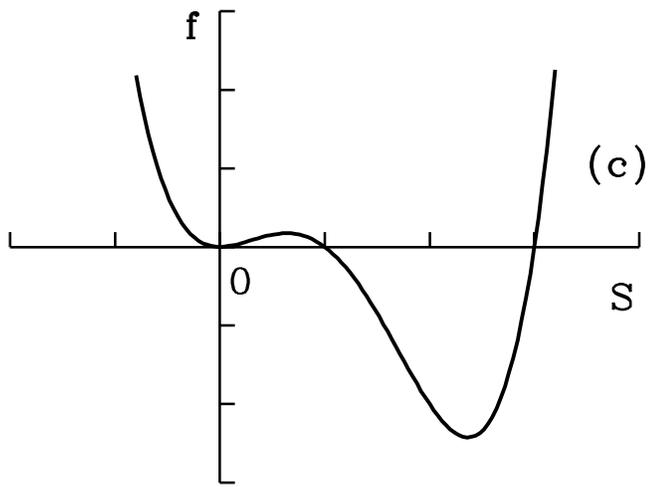
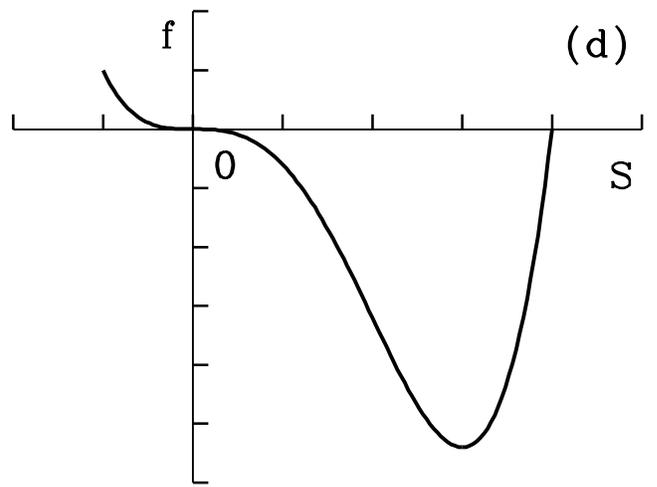
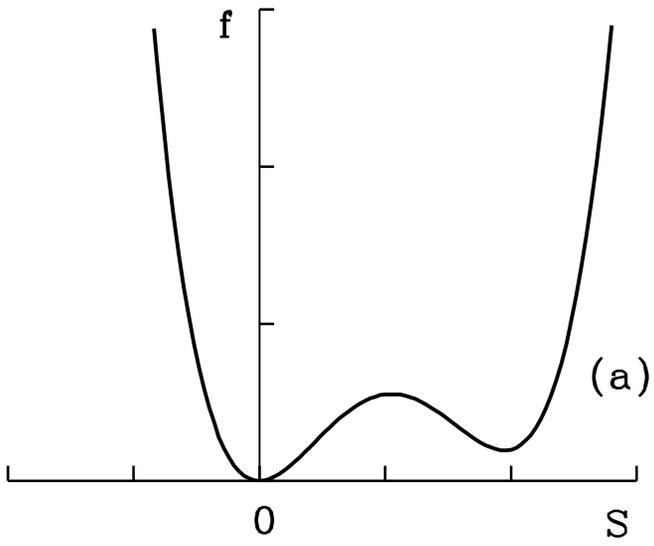
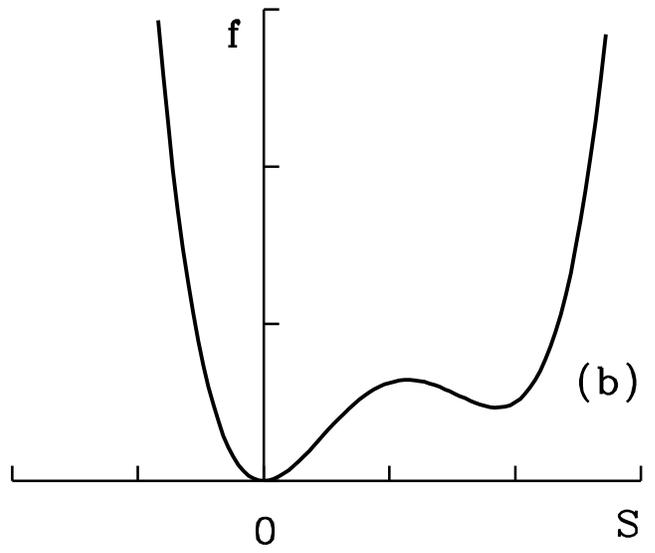
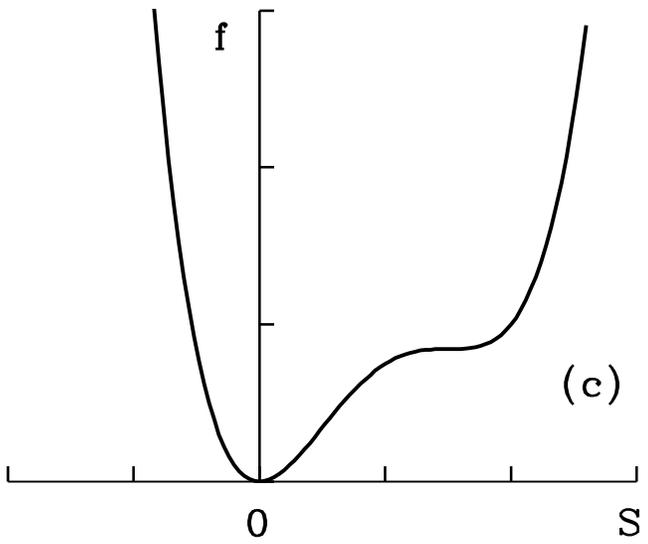
\end{document}